\documentclass[%
 prl, twocolumn,
superscriptaddress,
 amsmath,amssymb,
 aps,
]{revtex4-1}

\usepackage{graphicx}
\usepackage{dcolumn}
\usepackage{bm}
\usepackage{xcolor}
\usepackage{xr}
\usepackage{bm}
\usepackage{amsmath}
\PassOptionsToPackage{hyphens}{url}
\usepackage{hyperref}
\usepackage{url}

\usepackage{xurl}

\usepackage{enumitem}
\usepackage{accents}
\usepackage{comment}
\hypersetup{colorlinks,linkcolor={blue!90!black},citecolor={blue!90!black},urlcolor={blue!90!black}}

\DeclareMathAlphabet\mathbfcal{OMS}{cmsy}{b}{n}

\usepackage{amsmath}
\usepackage{mathtools}
\usepackage{euscript}

\begin{document}

\title{Photometric Decision-Making During the Dawn Choruses of Cicadas}

\author{Rakesh Khanna A.}
\email{rakeshkhanna.a@gmail.com}
\affiliation{Network Centric Systems,
Bharat Electronics Limited,
Ghaziabad,
India}
\author{Raymond E. Goldstein}
  \email{R.E.Goldstein@damtp.cam.ac.uk}
  \affiliation{Department of Applied Mathematics and Theoretical 
Physics, Centre for Mathematical Sciences,\\ University of Cambridge, Wilberforce Road, Cambridge CB3 0WA, 
United Kingdom}
\author{Adriana I. Pesci}
\email{A.I.Pesci@damtp.cam.ac.uk}
\affiliation{Department of Applied Mathematics and Theoretical 
Physics, Centre for Mathematical Sciences,\\ University of Cambridge, Wilberforce Road, Cambridge CB3 0WA, 
United Kingdom}
\author{Nir S. Gov}
\email{nir.gov@weizmann.ac.il}
\affiliation{Department of Chemical and Biological Physics, Weizmann Institute of Science, Rehovot 7610001, 
Israel}
\affiliation{Department of Physiology, Development and Neuroscience, University of Cambridge, Cambridge CB2 3DY, United Kingdom}

\date{\today}

\begin{abstract}
We report the first quantitative study of the onset of dawn
choruses of cicadas in several natural habitats.  
A time-frequency analysis
of the acoustical signals is used to define an order 
parameter for the development of 
collective singing.  The ensemble of recordings
reveals that the chorus onset times accurately
track the changing sunrise times over the
course of many weeks, occurring within
civil twilight at a solar
elevation of $-3.8^\circ \pm 0.2^\circ$.
Despite day-to-day 
variations in the amplitude of fully developed
choruses, the order parameter data collapse to a common 
sigmoidal curve when
scaled by those amplitudes and shifted by the onset time, revealing
a characteristic rise time of $\sim 60\,$s for a chorus to reach 
saturation amplitude.  The results are used to obtain the 
cumulative distribution function of 
singing as a function of  
ground illumination, from which is obtained a generalized susceptibility which exhibits a narrow peak with a 
half-width of $\sim\! 12\%$.  The variance
of the order parameter exhibits a similar peak, suggesting that a generalized fluctuation-dissipation theorem holds for this system. 
A model of decision-making under ramps of a 
control parameter is developed and can achieve a quantitative match to the data.  It suggest that sharpness 
of the susceptibility peak reflects cooperative decision-making arising from acoustic communication.
\end{abstract}

\maketitle

Among the most familiar collective behaviors in the 
animal world are the choruses of birds and insects at 
dawn and dusk.  In the case of birds, there is a 
long history of quantitative observational studies dating back 
well over a century \cite{Witchell1896} noting how 
dawn choruses track the seasonally changing sunrise 
times \cite{Allard1930} and the physiology of birds 
\cite{Davis1958,Leopold1961,Thomas2001,Bruni2014},  and, more recently, quantifying spatio-temporal aspects of 
multi-species choruses \cite{Farina2015}.
Many hypotheses have been advanced to explain {\it why} 
birds engage in choruses, with explanations 
ranging from diurnal variations in 
individual physiology to ones based on social functions
of the choruses \cite{Staicer1996,Gil2020}.
In addition, mechanisms that underlie the observed 
synchrony of singing have been
investigated both for birds and insects 
\cite{Greenfield1994,SiYuan2009,Sueur2002,Sheppard2020,Peleg2022,Ermentrout2022}.

Twilight bird choruses are but one example of collective
behavior in response to a changing
external cue.  A spectacular entomological example
on longer time scales 
is the vast swarms of periodical 
cicadas that emerge after a (prime) number of
years spent developing in underground burrows 
as the soil warms in springtime \cite{Simon2022}.
Seminal work \cite{Heath1968} showed that emergence is
associated with the soil temperature \textemdash and cicada body temperature\textemdash passing through a 
fairly well-defined threshold value, as evidenced by
the staggered emergence over several weeks of subpopulations in markedly different
microclimates (sunny south-facing hilltops,
shaded valleys).  

On closer inspection \cite{Goldstein2024} 
the problem of decision-making under such ramps in an external stimulus 
is complicated by the significant local
variations in soil temperature experienced by 
underground cicada nymphs even within a 
particular microclimate due to different burrowing depths and local insolation.
It was suggested \cite{Goldstein2024} that
communication between underground nymphs, most
likely by acoustic signaling, could overcome
this environmental noise and lead to larger, 
more coherent swarms than those that would be
produced by individuals responding only to
their perceived thermal environment.

Such considerations suggest that these kinds of 
collective decision-making problems conform to a 
conceptual picture advanced in the context of human 
decision-making \cite{Michard2005,Bouchaud2013}. There, 
each individual is aware of ``public information" that 
is known to all, perhaps with additive noise, and is coupled 
to a set of near 
neighbors who are also in the process of 
decision-making.  An individual's decision that a 
threshold in the public information has been crossed is 
thus affected by its neighbors' opinions.   These 
ideas naturally lead to theories based on the
random-field Ising model in which a uniform external 
field is slowly changed in time.  Examples 
of processes described by this approach range from coordinated applause at 
a concert to the mass selling of stocks.

\begin{figure*}[t]
\includegraphics[width=2.0\columnwidth]{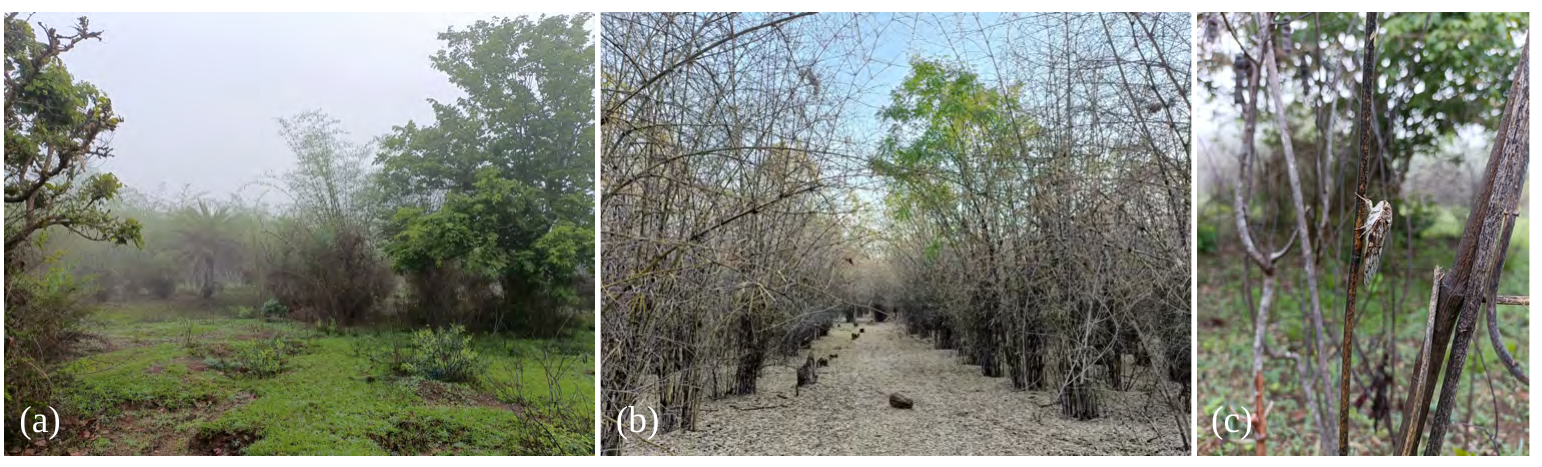}
\caption{Field observations. (a,b) Sites I and II, displaying range of flora. (c) Exemplar of \textit{P. capitata} 
on a tree.}
\label{fig1}
\end{figure*}

In light of these developments, avian and insectile choruses can serve
as paradigms with which to understand decision-making by populations
subjected to slowly changing cues.  While the vast body of prior work on
such choruses mentioned above has almost exclusively focused on 
defining only the apparent {\it start} of collective behavior, in reality the
choruses grow in amplitude over a finite time scale on the order of minutes.
In this paper, in contrast, we focus on the
detailed temporal development of the
dawn choruses of cicadas with the goal of quantifying how the 
synchronous singing develops in response to 
changing light levels.  This is done by
extracting from the acoustic signals an order
parameter for the amplitude of the choruses. 
We find first that the
dawn choruses on clear days commence at a sharply defined value of the pre-dawn 
solar elevation, which corresponds to a 
critical ground illumination level $I$, 
and that, in fact, 
on cloudy days the onset is delayed.
Secondly, the daily chorus amplitudes are 
found to grow sigmoidally as a function of light intensity and are self-similar.  
Third, we extract from the chorus amplitudes
a generalized susceptibility 
$\chi(I)$ of singing to light
and show
that $\chi(I)$ is sharply-peaked around a 
critical intensity $I_c$.  The fluctuations around the average $C$ are also found to peak at $I_c$, suggesting the existence of a generalized fluctuation-dissipation theorem at work. Finally, we develop a mathematical model for decision-making that suggests that this sharpness arises from 
collective effects.

Recordings of the choruses produced by the species \textit{Platypleura capitata} 
\cite{Pcap}
were obtained at two distinct
sites near Bangalore, India on multiple days in April and May of 2023, as indicated in Table \ref{tab1}.
Site I is a shrubland with scattered grasses 
and Site II is a 
bamboo forest.  
Figure \ref{fig1} displays photographs of the terrain and a typical example of {\it P. capitata}.

\begin{table}[t]
\caption{\label{tab1}%
Recordings locations and dates.}
\begin{ruledtabular}
\begin{tabular}{lll}
\textrm{Site}&
\textrm{Coordinates}&
Acquisition Dates\\
\colrule
I  & $13^{\circ} 9'18.82"$N, $77^\circ 28'18.04"$E & 24 April-1 May, 2023 \\
II & $13^{\circ} 9'28.63"$N, $77^\circ 29'48.99"$E & 15-20 May, 2023\\
\end{tabular}
\end{ruledtabular}
\end{table}

Stereo recordings were obtained at a sampling rate of $24\,$kHz using an
off-the-shelf recorder whose Nyquist 
frequency of $12\,$kHz is approximately 
twice the dominant frequencies found in the choruses (see below), assuring that they are
accurately captured.  The recorder was
mounted $\sim\! 1\,$m above the ground on 
a tree branch and remained in a single place for $8$ continuous days of recording at
each site, broken up into consecutive time-stamped $4$ hour recordings that were
spliced together for analysis.
Signals were analyzed with built-in and bespoke signal processing algorithms in 
Matlab.

\begin{figure}[b]
\includegraphics[width=1.0\columnwidth]{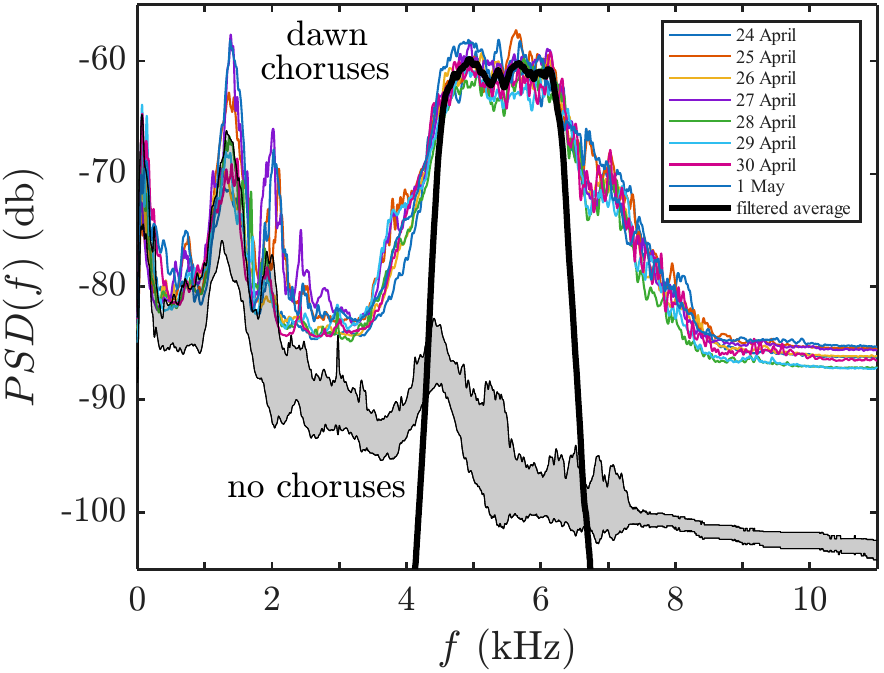}
\caption{Power spectra of dawn choruses at location I. Colored thin lines are raw power spectral densities of the $8$ data sets, while
heavy black line is the average of the $8$ bandpass-filtered signals.  Shaded gray region shows mean $\pm$ standard deviation for $8$ quiet periods after the choruses.}
\label{fig2}
\end{figure}

\begin{figure*}[t]
\includegraphics[width=1.8\columnwidth]{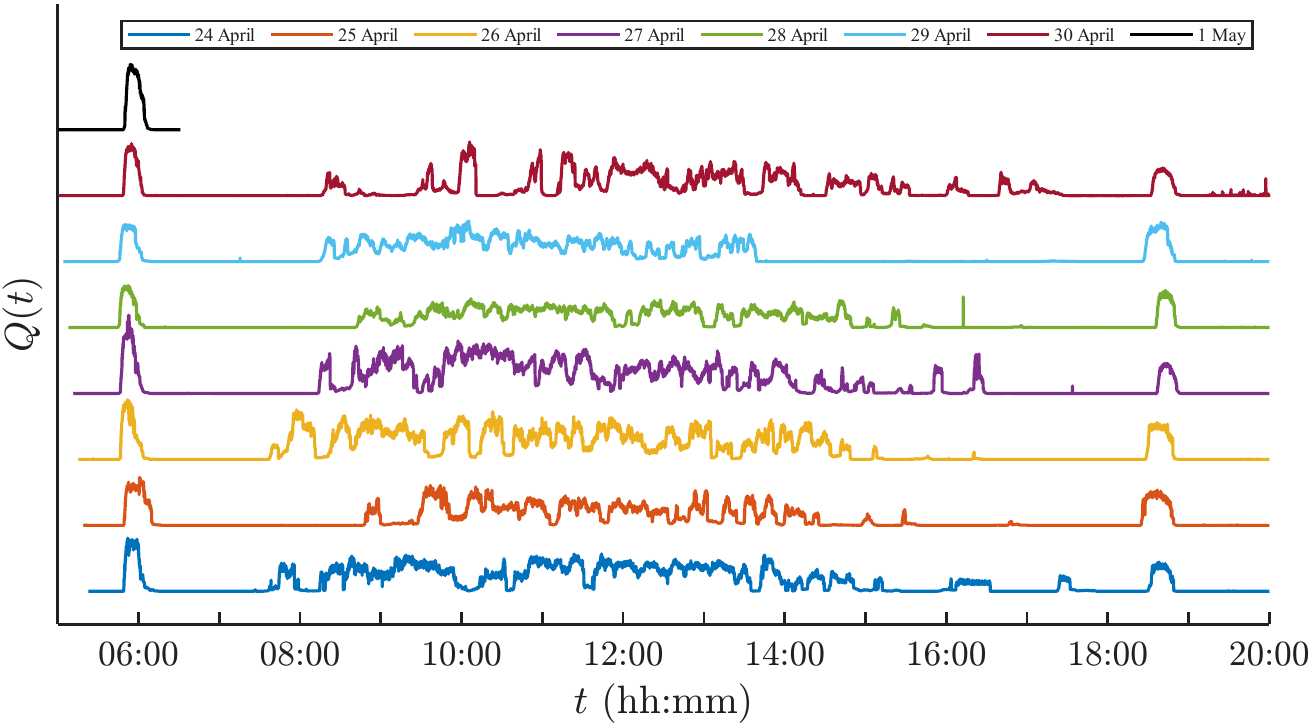}
\caption{Timeline of daily cicada choruses 
at location I.}
\label{fig3}
\end{figure*}

The recorded signals $S(t)$ are composed of the episodic cicada choruses superimposed on background noise arising from wind, other insects and distant environmental sounds.
A spectral analysis of the $8$ dawn choruses at location I, shown in Fig. \ref{fig2}, reveals structure at frequencies below $\sim 4\,$kHz whose amplitude varies considerably from day to day, and a highly reproducible peak above that
frequency, with a clear flat maximum extending
from a lower frequency of $f_l=4.3\,$kHz to
an upper frequency $f_u=6.3\,$kHz, the latter
being well below the Nyquist frequency.  
By analyzing segments of the recordings after
the end of each dawn chorus we obtained the
power spectra in the absence of singing.
Figure \ref{fig2} shows these data as bands of
$\pm$ one standard deviation around the mean.
It is clear from this that the background
noise is three orders of magnitude below
the main signal.
Based on this observation, we chose for all subsequent analysis to consider the filtered signal
$\tilde{S}(t)$ obtained by passing $S(t)$ through a $20^{\rm th}$ order Butterworth 
bandpass filter with lower and upper limits $f_l$ and $f_u$, yielding the truncated average
spectrum shown in Fig. 
\ref{fig2}.

There is a vast separation of time scales between 
the sub-millisecond period associated with the 
dominant chorus pitch and the many seconds over 
which the amplitude of the chorus develops to a 
quasi-steady value.  Thus, we can obtain a time-dependent measure of the amplitude of a chorus
by dividing up
the timeline of the signal into intervals of 
duration $T$ (in practice, we use $T=5\,$s) and computing
within each interval the 
power $Q(t)$ in the 
filtered signal $\tilde{S}(t)$ as
\begin{equation}
    Q(t)=T^{-1}\int_{t-T/2}^{t+T/2}dt'\,\tilde{S}(t')^2,
\end{equation}
which is a simple implementation of a time-frequency analysis \cite{Cohen1989}. For the pure tone $S=a\sin(2\pi f_0 t)$ 
with $f_0$ lying inside the bandpass window, then $Q=a^2/2$, and
$Q^{1/2}\propto a$ is a convenient 
measure of the
sound amplitude.  We thus define
the order parameter as
$A(t)=Q^{1/2}(t)$.

Using this procedure, Fig. \ref{fig3} shows $Q(t)$
of the daily choruses at site I.  Viewed on this coarse time scale, 
both the dawn and dusk choruses exhibit
highly reproducible onset times and more varied termination times. The mid-day 
choruses are highly variable in detail from day to 
day, although they are generally most noticeable 
during
the period $08\colon\! 00-16\colon\! 00$.

\begin{figure*}[t]
\includegraphics[width=2.0\columnwidth]{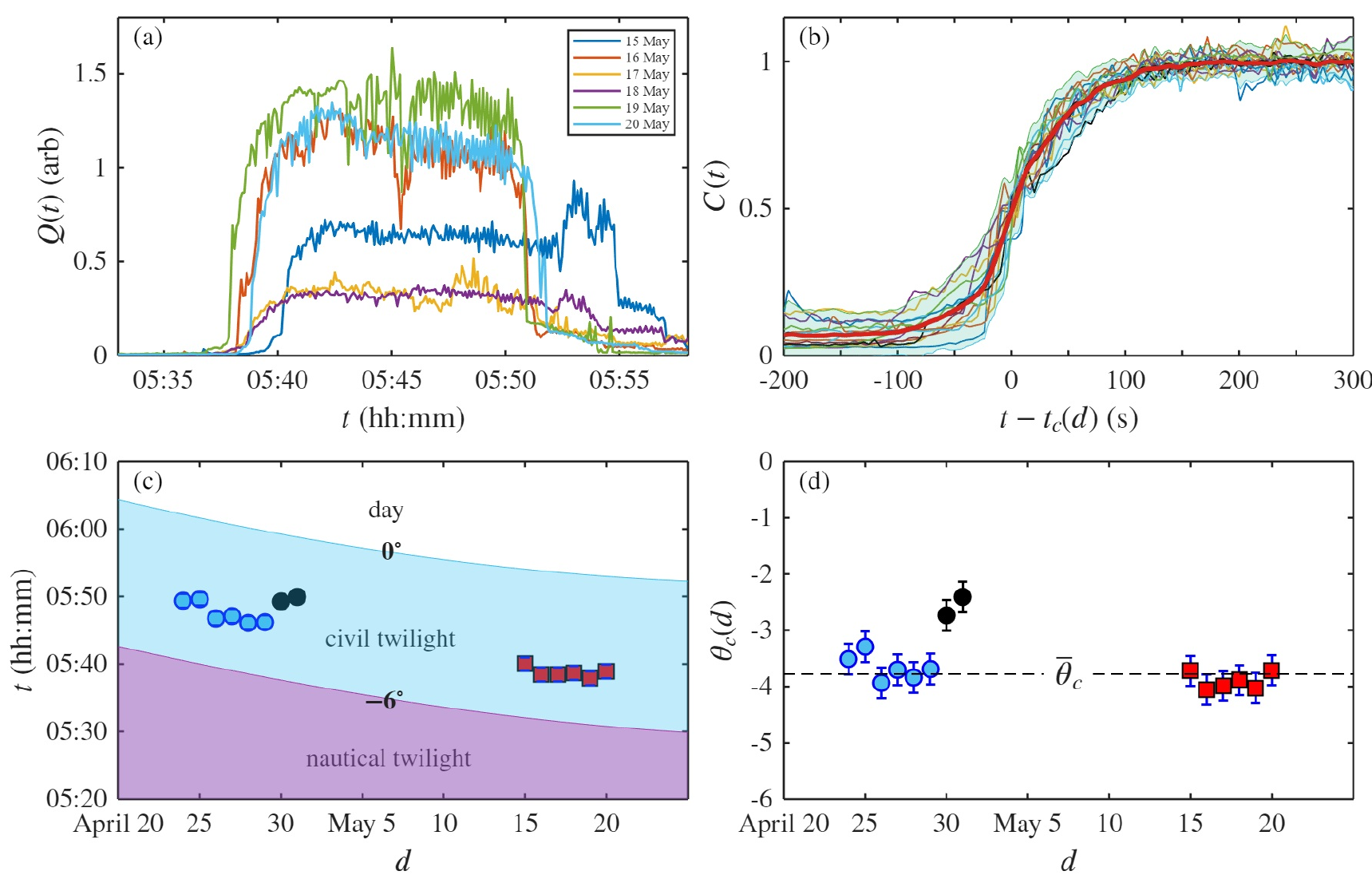}
\caption{Dawn choruses. (a) Magnified 
view of dawn chorus power $Q(t)$ at site II. 
(b) Chorus order parameter $C(t)$ near 
onset, for sites I and II, showing 
data collapse.  Heavy 
red line is mean, while shaded area represents $95\%$ 
confidence intervals. 
(c) Chorus onset times relative to civil twilight (solar elevation of $-6^\circ$)
and sunrise ($0^\circ$) for sites I (circles) and II (squares).  
For site I, light blue circles are for fair weather, black for
cloudy days. Smooth boundaries 
between the different periods are cubic interpolants of 
public data \cite{sunrisesunset}. 
(d) Solar elevation at chorus onset versus day. 
Average $\bar{\theta}_c$ excludes two 
cloudy days.}
\label{fig4}
\end{figure*}

In this paper we focus on the dawn choruses.
Figure \ref{fig4}(a) gives a magnified view of
the amplitude $A(t)$ as a function of time at location II, 
illustrating that the start of each chorus exhibits
a similar functional form, albeit with different
saturating values $A_{\rm max}(d)$ on each day $d$.  
The same behavior is found in the dawn choruses at site I.
Given 
these features,
it is natural to ask if the data can be 
collapsed under suitable scalings.  
For data from both sites I and II, 
Fig. \ref{fig4}(b) illustrates a test using what we term the 
\textit{chorus order parameter} 
\begin{equation}
    C(t)=\frac{A(t-t_c(d))}{A_{\rm max}(d)},
\end{equation}
where $t_c(d)$ is a day-dependent
shift termed the \textit{chorus onset time}, defined such that $C(t_c(d))=1/2$.
We see from the figure that the quality of the data
collapse is good, encompassing distinct natural locations 
several weeks apart.  A simple parameterization of the
sigmoidal shape is $C(t)=(1+\tanh(t/\tau))/2$, from which we obtain the 
estimate $\tau \sim 60\,$s for the characteristic
rise time of a chorus.  This time is well-resolved 
given that
it is an order of magnitude 
larger than the sampling window $T$. 
An interesting feature to which we 
return below is the markedly larger day-to-day fluctuations
around the mean near $t_c$.

We next examine how the chorus onset times $t_c(d)$ 
relate to the standard pre-dawn periods that are defined by the 
ranges of (negative) solar elevation angle $\theta$,
known as astronomical twilight ($-18^\circ \le \theta < 
-12^\circ$), nautical twilight ($-12^\circ\le \theta 
< -6^\circ$) and civil twilight ($\-6^\circ \le 
\theta < 0^\circ$).  
Data on the beginning and end of each of these
periods is readily available \cite{sunrisesunset}.
Figure \ref{fig4}(c) shows that 
$t_c(d)$ tracks the 
civil twilight/nautical twilight boundary 
$t_{CT}(d)$.  This is re-expressed in Fig. 
\ref{fig4}(d) as the solar elevation 
$\theta_c(d)$ at onset, computed (in degrees) as
\begin{equation}
    \theta_c(d)=-6\frac{t_R(d)-t_c(d)}{t_R(d)-t_{CT}(d)},
\end{equation}
where $t_R(d)$ is the sunrise time. 
Examination of historical records on cloud cover
\cite{cloud_data} shows that the two final days of the
data set at site I were significantly cloudy 
and preceded by heavy rain and thunderstorms, 
as verified by the audio recordings.  This strongly
suggests that the delay in the choruses is due to a lower light
intensity due to the cloud cover.
The records also show several days in which, 
while the weather was clear in the period 
leading up to the chorus, rain had occurred many 
hours earlier, leading to a noticeable increase 
in humidity and decrease in temperature.  
Nevertheless, the choruses on those clear days 
conformed to the general pattern, indicating 
that temperature and humidity do not play a role 
in the chorus onset. Excluding the data from site I on those cloudy days,  
the remaining data cluster very well
around the value $\bar{\theta}_c\simeq -3.8\pm 0.2^\circ$.

\begin{figure*}[t]
\includegraphics[width=1.98\columnwidth]{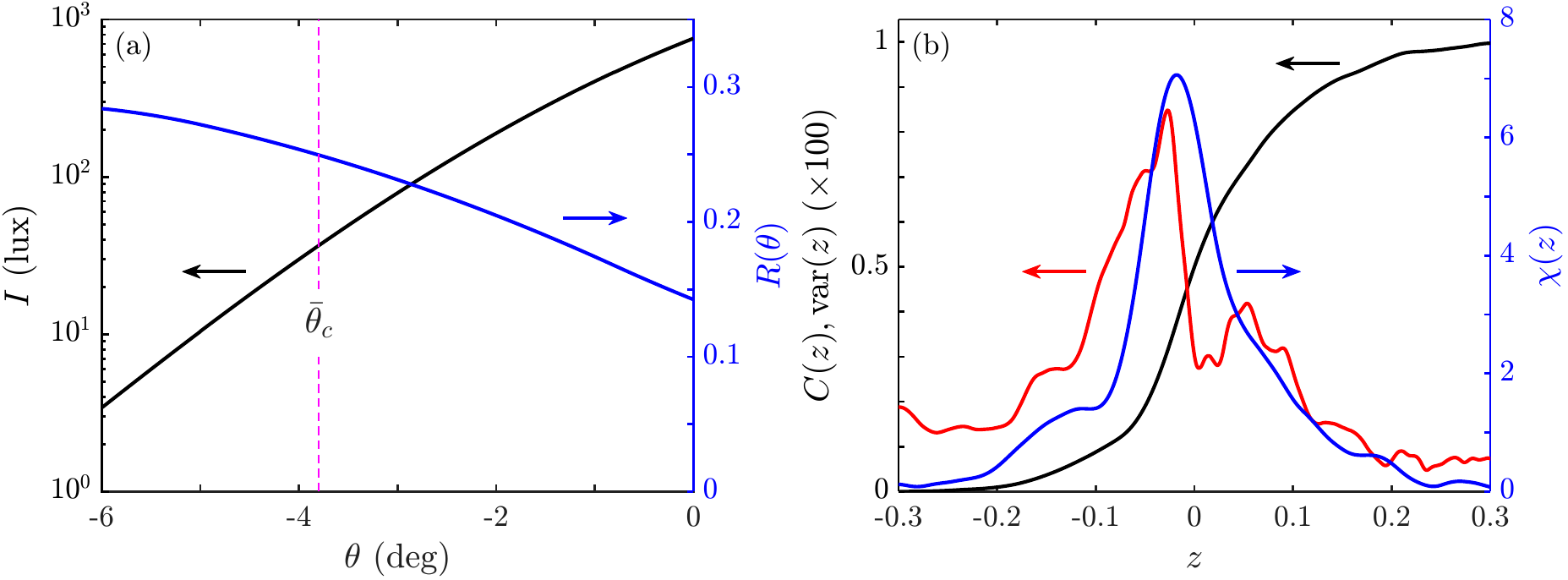}
\caption{Illumination and decision-making. (a)  Ground illumination $I$ as a function
of solar elevation $\theta$ (black, left axis) during 
civil twilight, using standard formulae \cite{almanac}. 
Right axis (blue) shows relative changes in illumination from \eqref{Rdef} during the period, using $\tau=60\,$s.   (b) Smoothed values of chorus order parameter $C(z)$ (black), variance ${\rm var}(z)$ of the
order parameter (red) and 
generalized susceptibility $\chi(z)$ 
(blue)
as functions of the logarithm of normalized ground illumination (\ref{zdef}).}
\label{fig5}
\end{figure*}

These results indicate the high precision with which
the choruses are associated with a particular
solar elevation.  This precision can be 
re-expressed in terms of the solar illumination $I$ 
experienced by the cicadas.
Direct measurements of the prevailing illumination
with a hand-held light meter showed values of 
$4-6$ lux at the beginning of the
choruses, and we may appeal to standard 
astronomical observations \cite{almanac} to obtain
estimates during the 
entirety of civil twilight.
Figure \ref{fig5}(a) shows that the 
ground illumination $I(\theta)$ varies 
over $\sim 2.5$ orders of magnitude as the 
sun moves from $-6^\circ$ to $0^\circ$.

In view of the relatively short time scale
$\tau\sim 60\,$s over which cicada choruses 
develop, it is natural to estimate
the changes in $I$ during that period.
Earth's rotational frequency $\omega_E$ is
$360^\circ/24\,$h, or conveniently $(1/4)^\circ\,$/min.  Thus, in the 
time $\tau\sim 1\,$min associated with the onset of
the dawn chorus the Earth rotates through 
an angle $\delta\theta=\omega_E\tau\sim 
(1/4)^\circ$.  As the rotation is slow 
compared to the chorus onset time, 
the relative
change $R(\theta)$ in ground illumination during 
the rise of the chorus can be estimated as 
\begin{equation}
R(\theta)=\frac{I(\theta+\delta\theta)-I(\theta)}{I(\theta)}\simeq \omega_E\tau 
    \frac{1}{I(\theta)}\frac{dI(\theta)}{d\theta}.
    \label{Rdef}
\end{equation}
As shown in Fig. \ref{fig5}(a), $R(\theta)$
varies by less than a factor of $2$ during
civil twilight, with $R(\bar{\theta}_c)\simeq 0.25$.  
We conclude that under cloudless conditions and 
in the absence of any obstructing foliage 
the ground illumination would vary by
$\sim\!25\%$ during the rise time of the dawn chorus.  

A more detailed examination of the photometric response involves viewing the chorus order parameter $C(t)$ not as a function of time, but rather as a function of the (time-dependent) ground illumination $I$.  This is analogous to the way in which the changing time-dependent ground temperature for cicada emergences is the relevant parameter \cite{Heath1968,Goldstein2024}.  
As it is typical for visual systems to exhibit logarithmic light sensitivity (the Weber-Fechner law \cite{logsensing}), we show in Fig. \ref{fig5}(b) $C$ 
as a function of 
\begin{equation}
    z\equiv \log(I/I_c) 
    \label{zdef}
\end{equation}
smoothed by a Savitzky-Golay (SG) filter.  To make a thermodynamics 
analogy, we view $C(z)$ as an order parameter as a function of the variable $z$ and hence there is a
generalized susceptibility $\chi(z)$ defined as
\begin{equation}
\chi(z)=\frac{dC}{dz}.
\end{equation}
The SG-filtered 
$\chi(z)$ is shown in blue in Fig. 
\ref{fig5}(b).  The central peak of $\chi(z)$ is reasonably well approximated by a Gaussian, with a
standard deviation of $0.06$ on a logarithmic scale or $0.14$ on a linear scale, 
reinforcing the conclusion that the
singing decision is made with $\sim 25\%$ 
precision with respect to the changing
illumination.  

Also shown in Fig. \ref{fig5}(b) is the SG-filtered
variance of the order parameter $C(z)$.  As can be
seen in Fig. \ref{fig4}(b), the scale of the standard
deviation is $\sim 0.1$, so the variance is $\sim 1\%$.
While small, it shows a somewhat noisy, but clear peak near $z=0$, and
in fact closely parallels the susceptibility $\chi(z)$.
A variance that is proportional to the susceptibility 
is evidence of a generalized fluctuation-dissipation
theorem \cite{Sato}, where, rather than being derivable from statistical physics, the 
proportionality constant must be viewed as a distinct property of each given system. 

In developing a model for these observations, we begin by noting that the significant variations in the 
local illumination based on surrounding vegetation levels and daily cicada positions and 
orientations, is a strong indication that the 
existence of
such a precisely defined transition to the dawn choruses requires a 
group decision-making process.
The notion that groups of organisms can make more accurate decisions when acting collectively 
\cite{Rauhut2011} has been explored in 
contexts ranging from animal and insect locomotion \cite{Couzin2008}
to discrimination between possible nesting sites of ants \cite{Sasaki2013}, 
bacterial density determination in the 
process of quorum sensing \cite{MorenoGamez2023}, and insect clocks
\cite{Rivas2016}, often using a statistical physics approach based on spin models  \cite{Hartnett2016}.


\begin{figure*}[t]
\includegraphics[width=1.98\columnwidth]{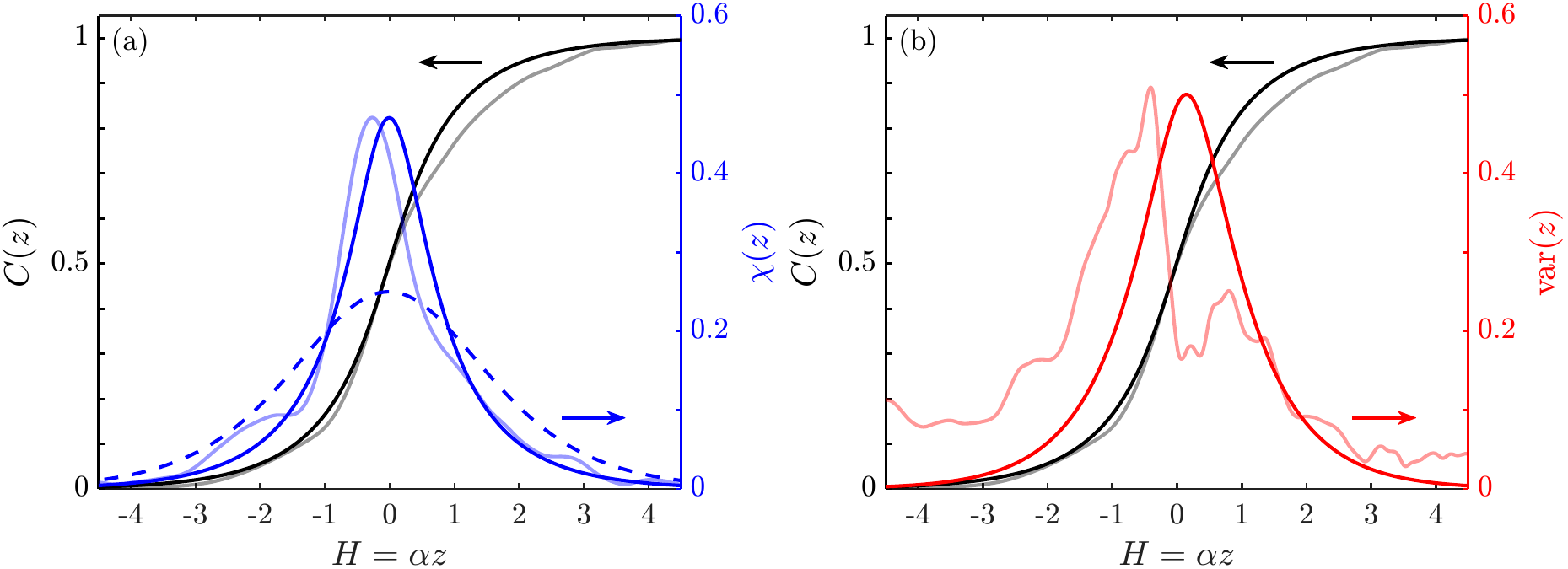}
\caption{Test of model for decision-making. 
(a) Chorus order parameter (black) and
susceptibility (blue) of optimum model compared with data from Fig. \ref{fig5}(b)
(grey, light blue).  Dashed blue line indicates
susceptibility for $J=0$.
(b) Comparison between model variance 
(red) and experimental variance (light red), the latter scaled by a factor of 
$60$ for the purposes of comparison.}
\label{fig6}
\end{figure*}

A simple model to describe the onset of singing, motivated by the spin description of animal activity and interactions \cite{Pink2018}, involves
assigning to each insect $i$ a  variable $n_i$, 
where $n_i=0(1)$ is the quiet (singing) state. The chorus order parameter is 
$C=\langle n\rangle$. 
The collection of variables is governed by
an energy $E$ with an external field $\tilde{H}$ and a coupling $\tilde{J}$,
\begin{equation}
    E=-\frac{\tilde{J}}{N}\sum_{i,j}n_in_j
    -\tilde{H}\sum_i n_i,
\end{equation}
where the light intensity, expressed through
the
variable $z$ in \eqref{zdef}, is represented 
by the external field $\tilde{H}\propto z$.
In this way, by itself, a sweep
from $\tilde{H}<0$ to $\tilde{H}>0$ would
result in the variables flipping from $0$ to
$1$.  Thus, $\tilde{H}$ corresponds to the
public information, and the infinite-range 
spin-spin
coupling $\tilde{J}>0$ (reflecting acoustic communication as in midge swarms \cite{Gorbonos}) 
fosters collective
behavior under the assumption that all cicadas behave identically.

In light of the long-ranged coupling between the spins and the fact that $N\gg 1$, the 
dynamics of transitions between states can
be described through a mean field 
approximation that takes the form 
\begin{equation}
\frac{dC}{dt}=(1-C)R_{0\to 1} -C R_{1\to 0},
\label{rate_eqn}
\end{equation}
where $R_{0\to 1}$ and $R_{1\to 0}$ are the relevant 
transition rates from quiet to singing 
and from singing to quiet. In the Glauber formalism these
rates are
\begin{equation}
    R_{0\to 1}=\frac{k}{1+e^{-JC-H}}, \ \ \ \ 
    R_{1\to 0}=\frac{k}{1+e^{JC+H}},
\end{equation}
where $k^{-1}$ is a decision time of a cicada.
Here, $J$ and $H$ are dimensionless variables
corresponding to $\tilde{J}$ and $\tilde{H}$
scaled by some effective thermal energy.
For $J=H=0$ the two transition rates
are both $k/2$.  For $J=0$ the transition
to singing is enhanced (and the transition
to quiet is diminished) for increasing positive $H$.  Similarly, for $J>0$, 
the larger
the order parameter $C$ the more
the transition to singing increases;
this is the collective effect on decision-making.

When $k\tau\gg 1$
we may use the quasi-static approximation $dC/dt=0$,
giving a self-consistency condition
on $C$ like that in the theory of 
ferromagnetism,
\begin{equation}
    C=\frac{1}{2}\left[1+\tanh\left(\left(JC+H\right)/2\right)\right].
    \label{Cself}
\end{equation}
When $J=0$, \eqref{Cself} yields
the sigmoidal curve 
$C=(1/2)(1+\tanh(H/2))$; increasing
positive $J$ produces a more pronounced sigmoid and larger susceptibility.  
Within this model, 
the variance of the order parameter can be 
expressed in a form similar to the rate equation
itself \eqref{rate_eqn}, as
\begin{equation}
{\rm var}(C)\propto (1-C)R_{0\to 1}+C R_{1\to 0}.
\end{equation}

In testing whether the solutions to \eqref{Cself}
are consistent with the data in Fig. \ref{fig5}(b), we must recognize that if $H=\alpha z$
(i.e., proportional rather than strictly equal),  
then $\alpha$ is to be determined along with $J$.   Numerical comparison between the
field data and the model shows that there is an approximately linear locus in the 
$\alpha-J$ plane along which reasonable fits
can be obtained, and that along this
locus there is a minimum in the 
total squared deviation between the 
data and the model (chi-squared) 
at $\alpha\simeq 15$ and $J\simeq 1.88$.
A comparison between the data and the 
optimum of this model is shown in Fig. \ref{fig6}(a), where we see very good agreement with $C(z)$ and $\chi(z)$. For comparison, the
result with $J=0$ reveals a considerably wider
and lower peak in the susceptibility. 
Figure \ref{fig6}(b) shows that an appropriately scaling of the experimental variance qualitatively matches the theoretical variance.
The fact that both 
$H$ and $J$ of the optimum model are the same order of 
magnitude indicates is 
support for the hypothesis that
decision-making in the dawn choruses
involves a synergy between 
external cues and collective 
effects.  In particular, the intermediate value of $J$ strikes a balance between on the one hand sharpening the response to changing light levels, giving rise to a rapid chorus development that may be important in such functions as mate attraction 
\cite{Hartbauer2016,Beauchamp2024}, and on the other avoiding spontaneous chorus development not strongly correlated with light levels.

In this paper we have presented a
framework for the analysis of data on collective behavior of decision-making groups insects, using the example of dawn choruses of cicadas. 
Our work highlights the balance that animals need to reach between individual sensory information (light level) and information shared between individuals (acoustic) in order to optimize their decision-making process. This is a process that occurs generally across collective animal behavior.
It is natural to ask whether the methods described here can be applied to a wider group of animals responding to various environmental cues.  Further tests of the hypothesis that decision-making in the dawn choruses involves communication between equivalent insects could involve 
longer-term field studies in additional natural habitats, and also
interventions to compare the behavior of isolated individuals to that of the group.

\begin{acknowledgments}
We thank Nelson Pesci for inspiring discussions and K. Leptos for helpful comments on the manuscript. This work was supported in part by the 
Complex Systems Fund at the University of Cambridge (R.E.G. and A.I.P.).  N.S.G. 
is supported by the Lee and William Abramowitz Professorial Chair of Biophysics  (Weizmann Institute) with additional support from a Royal Society Wolfson Visiting Fellowship.
\end{acknowledgments}

The data that support the findings of this article are openly available \cite{Zenodo}.


\end{document}